# Dilemmadata: On the Interoperability of Heterogeneous Roman Numeral Datasets

Johannes Hentschel[1], Emmanouil Karystinaios[2], Gerhard Widmer[2], Markus Neuwirth[1]

[1]Linz Computational Music Analysis Research Group (LCMA), Anton Bruckner University, Austria

[2]Institute of Computational Perception (CP), Johannes Kepler University Linz, Austria

<johannes.hentschel@bruckneruni.at>

## Introduction

The boom of deep learning and big data approaches in recent years has heavily impacted many fields, including digital musicology. Recent efforts have brought forth large-scale datasets of harmony-annotated scores enabling longitudinal studies and machine-learning approaches, notably the AugmentedNet Dataset (AND) (Nápoles López et al., 2021) and the Distant Listening Corpus (DLC) (Hentschel et al., 2025). Although both encode human expert analyses in the form of Roman numeral labels, in practice however, harmonising them into a single training and research corpus proved cumbersome and error-prone due to disparate encoding paradigms. We present the outcome of this effort: `dilemmadata`, the largest homogeneous Roman numeral dataset available to date. The name reflects the difficult yet unavoidable decisions necessitated by the reconciliation of such carefully crafted label sets. In Section 1, we will briefly introduce the source datasets and their underlying encoding paradigms, before discussing our solution to this reconciliation challenge (Section 2) and the learnings for the music encoding community (Section 3).

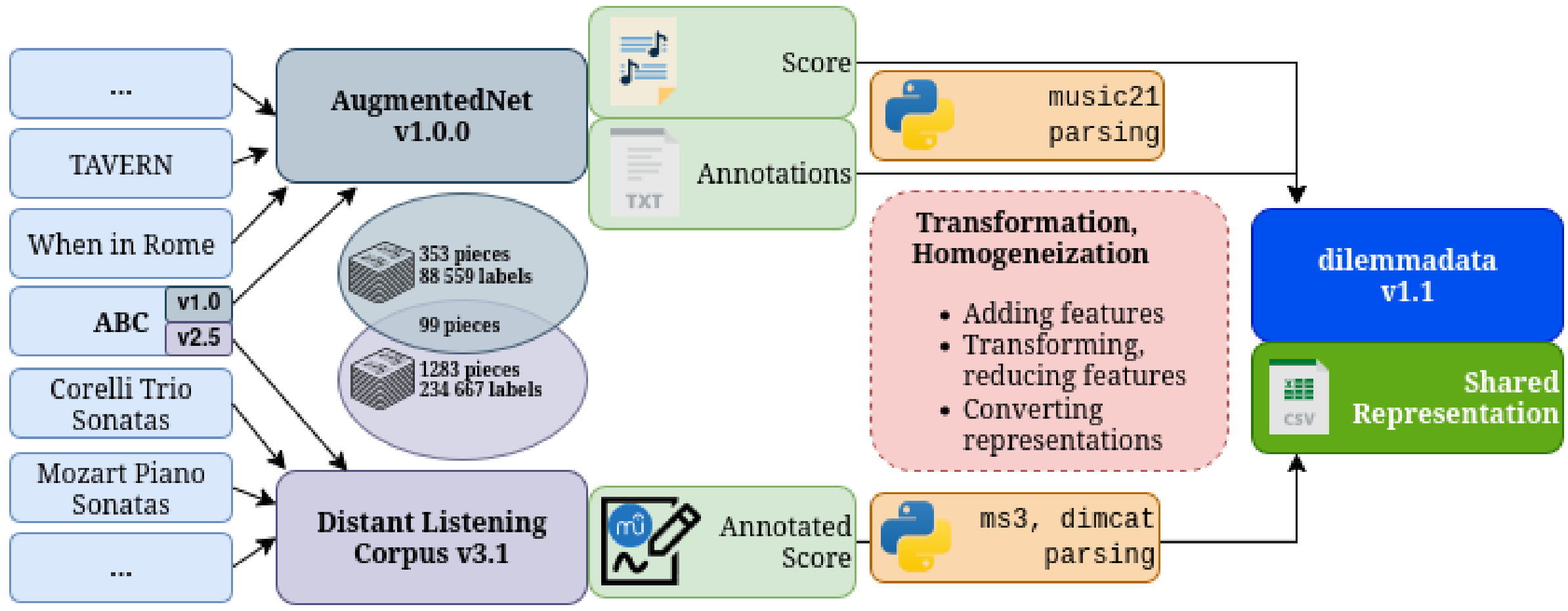

**Figure 1.** Data processing pipeline from left to right. The left side depicts the AugmentedNet Dataset and the Distant Listening Corpus as meta-corpora comprising multiple sub-corpora in their own right, one of them identical: the Annotated Beethoven Corpus, ABC, accounting for 70 of the 99 overlapping pieces (remaining overlaps stem from distinct sources and provide concurrent analyses). Each dataset's music representations (green rectangles) are processed with their respective parsing libraries (orange rectangles) while applying multiple transformations (red rectangle). The pipeline yields our aligned and homogenised dataset.

# 1 Datasets

## 1.1 The AugmentedNet Dataset (AND)

AugmentedNet[1] is an automatic Roman numeral analysis neural network developed by Néstor Nápoles López as part of his PhD research (Nápoles López, 2022). The accompanying dataset comprises 353 distinct pieces of tonal music, each annotated at the segment level in the RomanText ('.rntxt') format–a human- and machine-readable plain-text standard that encodes chord root, quality, inversion, added intervals, and analyst notes, with full support for bidirectional conversion in `music21` (Cuthbert & Ariza, 2010; Tymoczko et al., 2019). Source scores originate from diverse formats (MusicXML, Humdrum, ABC), and export as TSV slice-segmented tables, including fields for onset, duration, beat position, and Roman numeral labels. Dataset construction aggregates analyses from six publicly available corpora–ABC, BPS, HaydnSun, TAVERN, WiR and WTC (Devaney et al., 2015; Nápoles López, 2017; Chen & Su, 2018; Gotham et al., 2023). Many works on automated Roman numeral analysis also train on and compare to the AND, effectively establishing it as a standard baseline for Roman numeral prediction models (Karystinaios & Widmer, 2023; Sailor, 2024).

## 1.2 The Distant Listening Corpus (DLC)

The DLC[2] provides 1,283 pieces from four centuries, symbolically encoded in MuseScore's MSCX format and enriched with embedded DCML harmony annotations, including Roman numerals, key, phrase, and cadence labels, directly integrated as textual objects within the score files (Hentschel et al., 2025). These annotated scores have been collaboratively created and reviewed by a team of trained music theorists using a distributed curation pipeline on GitHub (Hentschel, Moss, et al., 2021), ensuring notational well-formedness through automated `ms3` validations and inter-annotator consensus. Each sub-corpus, spanning more than 40 thematic collections from J.S. Bach's English Suites to N. Medtner's Tales, is released as both a GitHub repository and a frictionless data package (Fowler et al., 2018), with TSV exports of notes, harmonies, and markup available for downstream computational use. We removed 15 pieces overlapping with AND (see Section 2.3), leaving us with 1,268 annotated DLC pieces.

[1] https://github.com/napulen/AugmentedNet

[2] https://github.com/DCMLab/distant_listening_corpus

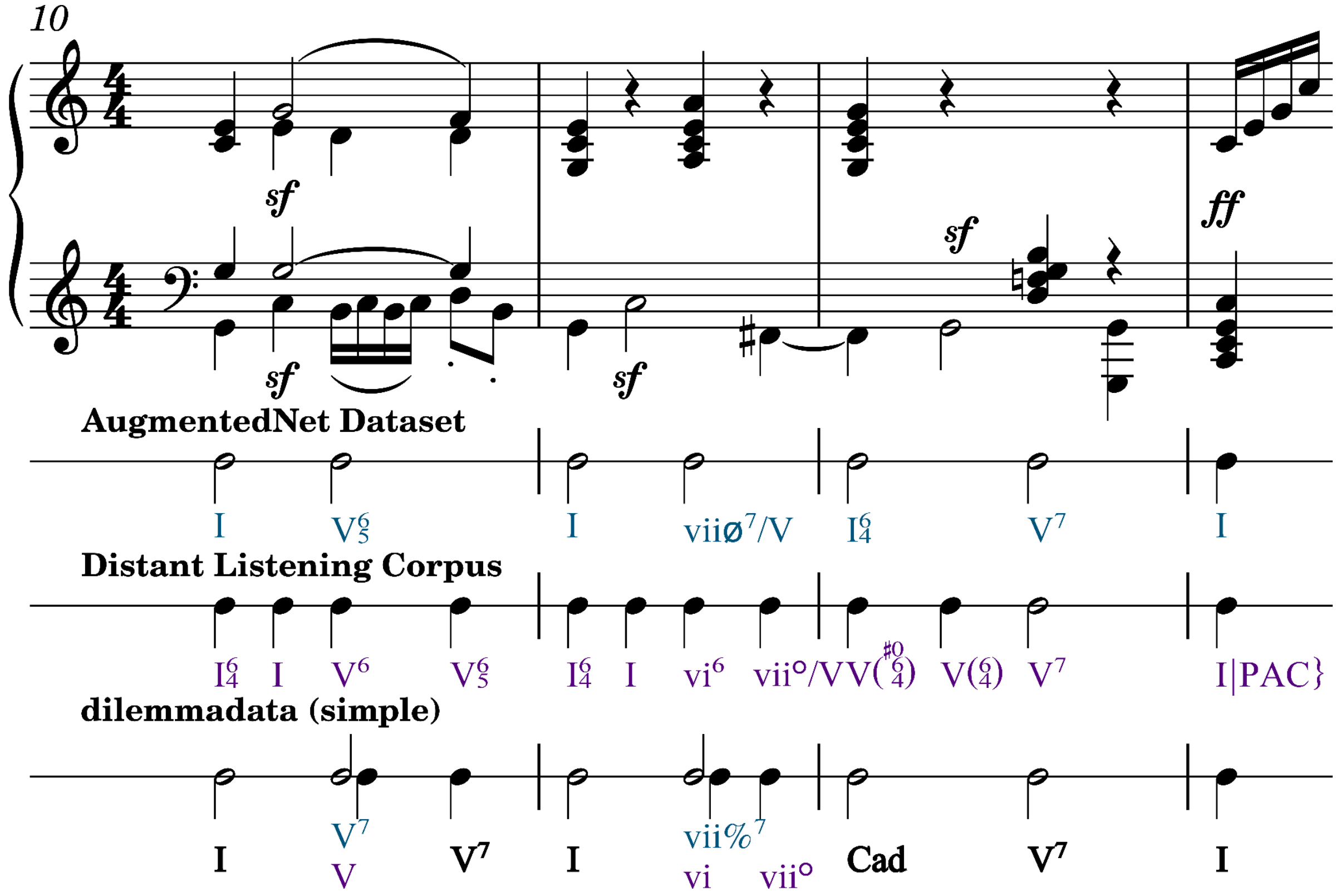

**Figure 2.** Measures 10-13 from the first movement of L. v. Beethoven's, Piano Sonata No. 3 in C major, Op.2, No. 3. The three annotation layers show labels and harmonic rhythm from the two source datasets and, in the bottom system, of our respective simplified representation without inversions and secondary keys (which are represented elsewhere in our dataset).

# 2 Dataset Alignment

## 2.1 Disparate encoding paradigms

The AND relies on the RomanText encoding standard and paradigm where Roman numeral analyses are stored as stand-off text files that can be paired with digital score encodings using `music21` (Cuthbert & Ariza, 2010; Tymoczko et al., 2019). DLC annotations, on the other hand, are encoded directly within MuseScore files, adhere to the DCML harmony annotation standard, and can be parsed and analysed using `ms3` and `dimcat` libraries (Neuwirth et al., 2018; Hentschel, Neuwirth, et al., 2021; Hentschel & Rohrmeier, 2023; Hentschel et al., 2023). As shown in Figure 1, we employed the respective Python libraries to transform the datasets into a shared CSV representation containing one row per note, including the pertinent annotation label. This involved adapting the original AugmentedNet processing pipeline, bypassing its creation of fixed-size, sixteenth-note slicing windows in favour of a note-wise representation (thereby expanding the dataset from roughly 100K annotated slices to over 750K individual annotated notes). For the DLC, we achieved the same representation by unfolding repeats and joining the tabular note and annotation data.

### 2.2 Disparate annotation standards

Figure 2 illustrates the primary discrepancies between the RomanText and the DCML harmony annotation standards. Between the two analyses, the one from the DLC has a perceptibly higher granularity and, thereby, a faster harmonic rhythm, attributable partly to the DCML standard's elaborate syntax for suspensions (expressed within parentheses). Consequently, the two standards yield analyses that frequently diverge in terms of chord types and which notes are to be considered as chord vs. non-chord tones. Whereas in RomanText, cadential 6/4 chords are annotated as either **I64** or **Cad64**, the DCML standard consistently encodes them as **V(64)** suspension chords, differentiating them from I64 inverted tonic harmonies (which have the same pitch content). The chord type syntax is slightly different (e.g. ø vs. **%** for the half-diminished seventh chord) and, complicating matters further, the AND assembles datasets that differ in their usage of the RomanText syntax.

We addressed these discrepancies by distributing the encoded information of each annotation label across dedicated feature columns such as root, bass, chord type, local key, etc. This approach enables users to flexibly choose between those features that are covered by both standards and those made available by the DCML standard exclusively (such as suspensions, pedal points, cadences, to name a few). This involved crafting shared vocabularies for the overlapping features, each followed by careful validation. For example, `dilemmadata` uses a harmonized chord type vocabulary which is the outcome of one-to-one mappings where possible, and manual adjustments (verified against the scores) where necessary. Importantly, we carefully replaced each occurrence of the cadential 6/4 chord, the seventh most frequent harmony, with the symbol **Cad**. Based on the harmonized features, we re-synthesized simplified chord labels (see Figure 2), offering a more general chord representation with a more homogeneous harmonic rhythm (when removing immediate duplicates).

### 2.3 Overlapping pieces

To reconcile AND and DLC, we compiled per-piece metadata for both collections, documenting available annotation features, provenance, and encoding formats. We then identified overlapping pieces by matching file and folder names, metadata fields, and, where necessary, through manual inspection of score content. This process yielded 99 pieces common to both datasets, of which 15 were removed because they appeared in the AND test set; the remaining 84 overlaps were retained as shared reference material. To support evaluation and downstream tasks, we also constructed a 20% test set of the DLC by selecting pieces on a per-sub-corpus basis, excluding collections too small for reliable sampling or too skewed towards a single mode.

## 3 Outcomes and Future Work

The final `dilemmadata` corpus comprises 1,621 pieces (353 from the AND plus 1,268 from the DLC) and over 2.8 million note-wise Roman numeral annotations, each enriched with precise onset, duration, beat-level features, chordal root, quality, inversion, extensions, and contextual metadata. Annotations carry provenance hashes and validity flags to facilitate selective filtering. All TSV files, metadata descriptors, and precomputed test splits are available for download[3]. For a full list of features, we refer our readers to the README.

Beyond the dataset itself, our contribution demonstrates that discrepancies between Roman numeral datasets produced under disparate paradigms extend far beyond mere syntax; they create severe impediments to integration and interoperability. The substantial effort required to carefully transform annotations for equivalence without distorting the original semantic intent underscores, in our view, the urgent need for a generalised data model capable of interfacing with the diverse harmony encoding standards currently in use (e.g., Temperley & Clercq, 2013; Cambouropoulos, 2016; White & Quinn, 2016; Chen & Su, 2018; Huron, 2020; Hentschel et al., 2022). Ultimately, we propose the overlapping `dilemma` pieces, annotated under both source paradigms, as a catalyst for replacing the prevailing reliance on objective "ground truth" in training scenarios, inviting the encoding community to discuss paths toward contingent, probability-based target labels that better reflect the inherently ambiguous nature of harmonic analysis.

## Acknowledgements

This research has been supported by the Swiss National Science Foundation (SNSF) through the project "Towards a Unified Model of Musical Form: Bridging Music Theory, Digital Corpus Research, and Computation" (grant no. 10000183; 2024-2028) and by the European Research Council (ERC) under the EU's Horizon 2020 research & innovation programme, grant agreement No. 101019375 (*Whither Music?*).

## Statement on the use of generative AI

We have employed a Large Language Model for re-formulating sentences from our original draft in a more concise and idiomatic way.

[3] https://zenodo.org/records/19661224